# Utilizing Transaction Prioritization to Enhance Confirmation Speed in the IOTA Network

seyyed ali aghamiri
School *of Electrical and Computer Engineering*
*University of Tehran*
Tehran, Iran
aliaghamiri@ut.ac.ir

Reza Sharifnia
School *of Electrical and Computer Engineering*
*University of Tehran*
Tehran, Iran
reza.sharifnia@ut.ac.ir

Ahmad Khonsari
School of Electrical and Computer Engineering
*University of Tehran*
Tehran, Iran
a_khonsari@ut.ac.ir

*Abstract*— With the rapid advancement of blockchain technology, a significant trend is the adoption of Directed Acyclic Graphs (DAGs) as an alternative to traditional chain-based architectures for organizing ledger records. Systems like IOTA, which are specially designed for the Internet of Things (IoT), leverage DAG-based architectures to achieve greater scalability by enabling multiple attachment points in the ledger for new transactions while allowing these transactions to be added to the network without incurring any fees. To determine these attachment points, many tip selection algorithms commonly employ specific strategies on the DAG ledger. Consequently, recent research has moved from chain-based blockchains to Directed Acyclic Graph (DAG)-based blockchains, aiming to reduce transaction confirmation times.

Prioritization is a strategy designed to enable faster confirmation of specific transactions. IoT applications can mark the priority level of essential transactions. Transaction prioritization is not considered in the IOTA network, which becomes especially important when network bandwidth is limited. In this paper, we propose an optimization framework designed to integrate a priority level for critical or high-priority IoT transactions within the IOTA network. We evaluate our system using fully based on the official IOTA GitHub repository, which employs the currently operational IOTA node software (Hornet version), as part of the Chrysalis update (1.5). The experimental results show that higher-priority transactions in the proposed algorithm reach final confirmation in less time compared to the original IOTA system.

## I. INTRODUCTION

The success of chain-based blockchain over the past years has demonstrated the real-world potential of blockchain technology. However, its limitations restrict its suitability as a universal cryptocurrency platform. A key drawback of chain-based blockchains is scalability, especially in the growing IoT industry, where a large number of transactions need to be published instantly [1]. Therefore, researchers are searching for alternatives to these blockchains. One of these alternatives is the Directed Acyclic Graph (DAG)-based distributed ledger for the Internet of Things (IoT). DAG-based blockchains leverage the Directed Acyclic Graph data structure as the primary storage framework, supporting concurrent transaction additions. Its intrinsic parallel data structure significantly accelerates block generation and offers notable performance advantages over traditional chain-based blockchains [1],[2],[3].

DAG-based blockchains have made significant strides in improving consensus scalability. Consensus protocols for DAG-based blockchains are classified into probabilistic and deterministic types. Probabilistic approaches, like IOTA [1], rely on transaction depth or cumulative weight but face latency and security issues. Deterministic protocols, such as BullShark [4], leverage Byzantine Fault Tolerant (BFT) consensus, decoupling transaction dissemination from ordering to achieve high throughput and fast confirmation.

In a DAG ledger, a vertex represents a message, and a directed edge signifies approval from one vertex to another. This structure offers several advantages, including multiple attachment points for new messages, lightweight consensus mechanisms, and the elimination of miner/transaction fees. [5], [6].

IOTA is an innovative cryptocurrency introduced in 2015, utilizing a quantum-resistant approach. It replaces the traditional blockchain with DAG-based distributed ledger technology, specifically designed for the IoT industry. The IOTA approach is ideal for IoT devices as it avoids the computationally intensive mining process, allowing it to operate on lightweight devices with limited power and memory [1].

The Tangle is the immutable data structure that stores transactions issued by IOTA nodes. Unlike traditional blockchains, transactions in the IOTA Tangle can be processed in parallel. A node, such as a computer or mobile device, issues and validates transactions, and each node maintains a copy of the Tangle. The security of the IOTA Tangle is ensured by two previously validated transactions in the network. Unconfirmed transactions are referred to as "tips" in the IOTA Tangle [7].

Prioritization is a strategy to identify critical transactions, such as those sent by IoT devices. This priority-based scheduling approach allows IoT applications to flag essential transactions. The use of priority level for necessary transactions is one of the essential parts in dag-based blockchains.

In this paper, we have implemented the original IOTA system and aimed to incorporate transaction prioritization for faster confirmation with minimal changes to the original system.

The remainder of the paper is organized as follows. Section II provides the background and related work. Section III gives priority-based tip selection method. Section IV describes experimental results of the PTSA, and Section V concludes the paper.

## II. BACKGROUND AND RELATED WORK

In this section, we first introduce some basic terminology used throughout this paper. Then, we provide a brief overview of related work on blockchain applications and DAG-based storage models.

A Directed Acyclic Graph (DAG) is a data structure made up of nodes linked by edges without forming any cycles, used to represent transactions and their dependencies within the Tangle. The Tangle, built on a DAG, stores blocks and their relationships. It serves as the foundational data structure of IOTA [7].

The tip selection algorithm plays a crucial role in a DAG-based system, determining which unconfirmed transactions should be validated by new transactions. In such systems, there is always a small chance that some valid transactions may never be approved. If a large number of transactions remain unapproved, the graph is considered to be unstable [1].

As illustrated in Fig. 1, new transactions added to the Tangle network are referred to as tips. These are transactions that have not yet been confirmed by any other transaction (highlighted in yellow).

Tip selection refers to the process of choosing prior transactions to reference in a new transaction, which enables it to link to the existing data structure. In IOTA, it is required that each transaction approves up to eight other transactions, while the tip selection strategy is left to the user, with a default option provided by Shimmer [7].

Confirmed transactions are those whose cumulative weight exceeds the confirmed cumulative weight threshold (depicted in pink). Cumulative weight is a method used to evaluate transactions, where each new transaction that references a previous one increases its cumulative weight. Transactions that have been validated by nodes but have not yet surpassed this threshold are considered unconfirmed transactions (depicted in blue). The first transaction in the network is referred to as Genesis (shown in green).

Xiao et al. [8] proposed precomputing the tip selection probability distribution to accelerate the process, introducing a new algorithm for burst message arrivals on edge nodes. In contrast to the weighted random walk approach, their solution focuses on precomputing the probability distribution for the DAG ledger, simplifying tip selection into a fast sampling task. The authors implemented and compared this approach with the random walk method, demonstrating its effectiveness in reducing edge node congestion and providing a new perspective for tip selection in DAG-based blockchain systems.

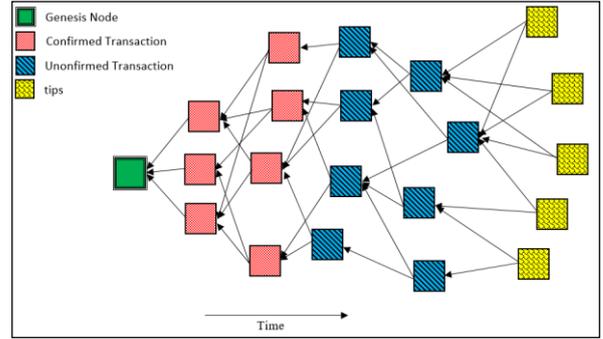

Fig. 1. The DAG structure of the IOTA Tangle

M. N. Halgamuge in [9] proposed a model to track newly arrived tips within a selected timeframe and calculate the transaction time before they are added to the Tangle network. In addition, the author introduce the Best Tip Selection Method (BTSM) to improve network security and incorporate priority levels for essential or prioritized IoT transactions. Halgamuge's time-based approach divides the Tangle into timeframes to select the most verified transactions, enhancing both security and scalability. This method also facilitates prioritization and emphasizes the timing of transactions.

## III. PRIORITY-BASED TIP SELECTION METHOD

In this section, we propose a priority-based tip selection method for IOTA system. Our goal is to expedite the final confirmation of high-priority transactions in the IOTA network. Additionally, transactions that remain unconfirmed for an extended period are designed to gradually gain higher priority.

Our priority-based tip selection method shown in Fig. 1. This method takes the DAG ledger structure as input and delivers two tips as output. In the proposed model, transaction verification incorporates the priority levels of transactions issued by IoT devices. Thus, the first step involves evaluating the transaction's priority within the DAG structure, and based on the count of unconfirmed high-priority transactions, the required number of transactions are selected to be confirmed as tips.

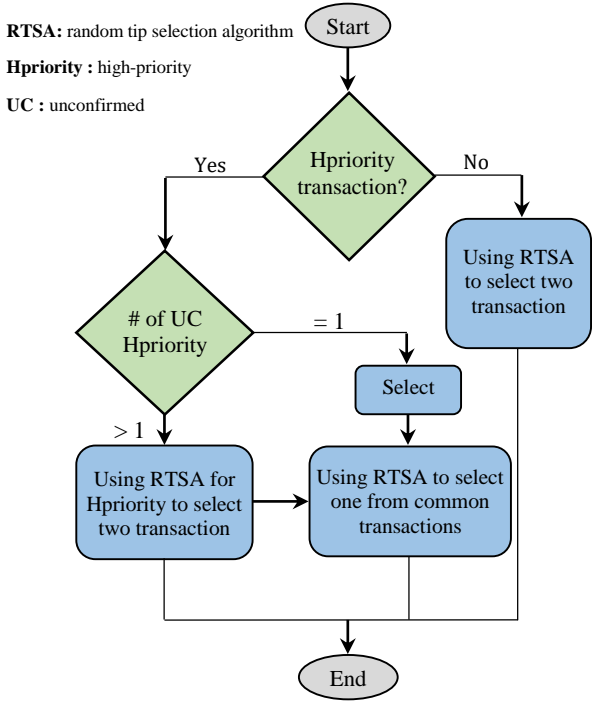

Fig. 1. The architecture of our priority-based tip selection method

Fig. 2 illustrates the PTSA algorithm, which is designed for tip selection while considering transaction priorities.

**Algorithm *PTSA*: Priority-base Tip Selection Algorithm**

**input**:
DAG-based distributed ledger structure
Priority of transactions
**output**:
  $Tips[1,2]$: Two new tips
1.   $p = 0$ // $p$ counts the number of priority nodes
2.   **for** each $x_i$ **do**   // $x_i$ are candidates for tip selection
3.       **if** $x_i$ has priority and not yet confirm **then**
4.           $p++$
5.       **end if**
6.   **end for**
7.   **if** $p = 0$ **then**
8.       calculate random tip selection method for common transaction and select two tips
9.   **else if** $p = 1$ **then**
10.      select the high-priority transaction and one common transaction
11.  **else**
12.      select two high-priority transaction and one common transaction
13.  **end if**

Fig. 2. PTSA algorithm for tip selection with transaction priorities

The for loop in the second line of the algorithm is employed to count the number of priority transactions that have not yet been confirmed (lines 2–6). Lines 7–8 address the absence of any priority transactions that are still unconfirmed. In this case, the random tip selection method is applied for common transaction and select two these transaction as tips. Lines 9–10 describe the scenario where there is only one priority transaction in the tip pool. In this case, in addition to selecting this transaction to continue the algorithm, another common transaction with the random tip selection method is also selected. When there are more than one high-priority transaction that are still unconfirmed, the algorithm uses random tip selection to select two high-priority transaction and one common transaction(lines 11–12). This additional common transaction selection ensures that non-priority transactions are not left waiting for final confirmation for an extended period. In any case, high-priority transacions are selected for confirmation faster than other transacions, and this additional selection does not disrupt the confirmation process for these transacions.

## IV. EXPERIMENTAL RESULTS

In this section, we present experimental results of our priority-based tip selection method for IOTA system. The prototype was deployed on a host machine featuring an Intel(R) Core(TM) i5-8250U CPU @ 1.60GHz 1.80GHz processor, 8GB of memory, and running the Ubuntu operating system. Our experiments analyze issues using the official IOTA private tangle framework, based entirely on the IOTA GitHub repository and the operational Hornet node software [11] .

We executed the use cases to investigate the issues outlined in Section III. Minor adjustments were made to the configuration of the IOTA private Tangle implementation to address these challenges. To enable prioritization in the IOTA tangle, the backend was modified by introducing a new feature called `priorityFlag`, which is assigned a value of 1 for high-priority transactions and 0 for all others. To prevent starvation in the selection of transactions as tips, transactions that have been in the network for a longer time without being confirmed gradually gain higher priority over time.

Fig. 3 illustrates the confirmation time of high-priority transactions in two IOTA tangles. One that utilizes transaction prioritization for confirmation and the other, the main IOTA system, which does not employ transaction prioritization. High-priority transactions are distinguished by a black color, and for better visibility, we have used numbers 1-5 to represent these transactions. As shown in this figure, higher-priority transactions in the proposed algorithm reach final confirmation faster than in the original IOTA system.

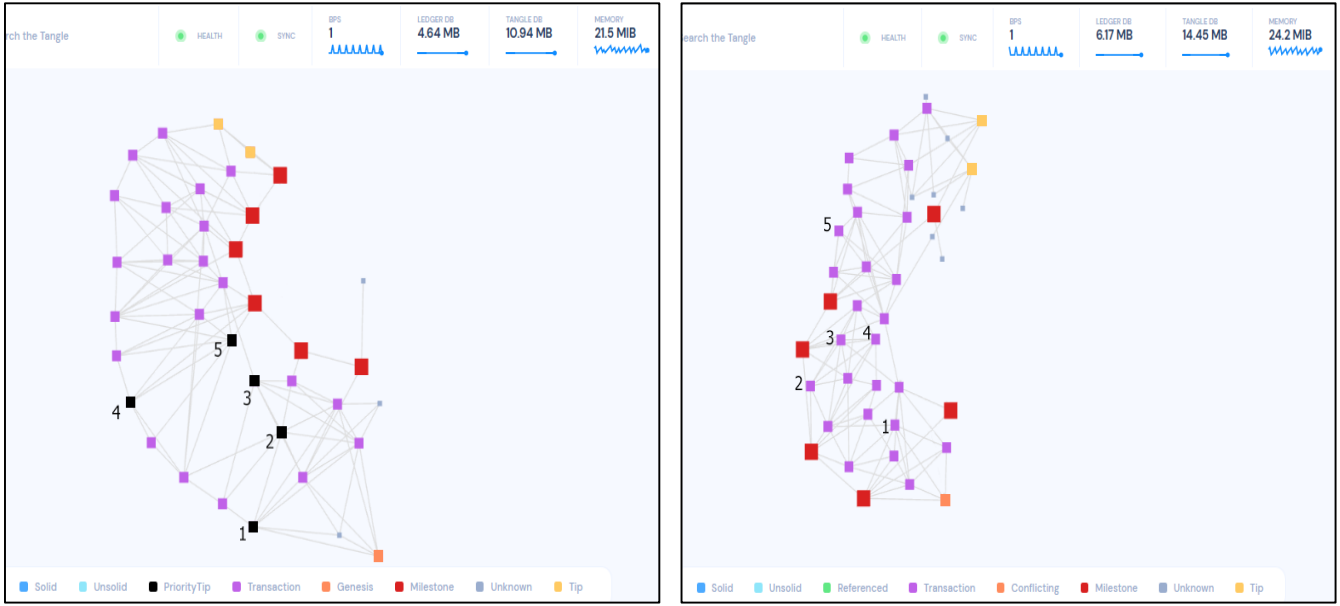

Fig. 3. Comparison of the final confirmation time of high-priority transactions in the IOTA system and the proposed model

## V. CONCLUSION

In recent years, blockchain technology has gained significant interest among researchers and are considered one of the most promising technologies. The growing number of users and connections demands higher scalability and improved throughput in blockchains. Currently, blockchains face the challenge of being decentralized, secure, and scalable simultaneously. To overcome these challenges and extend blockchains to the IoT, IOTA was introduced in 2015 to eliminates the limitations of traditional blockchains by using a Directed Acyclic Graph (DAG) as its underlying structure. However, in the IOTA system, transaction prioritization for faster confirmation has not been considered, and this lack of prioritization can lead to issues in certain cases.

In this paper, we propose a priority-based transaction system called PTSA, which utilizes the IOTA platform with minimal modifications to facilitate the final confirmation of high-priority transactions. Our experimental results show that the proposed system confirms higher-priority transactions faster than the original IOTA system, while ensuring that the remaining transactions do not become orphaned.